\documentclass[preprint,aps,tightenlines,nofootinbib]{revtex4}
\usepackage{graphicx}
\usepackage{bm}
\usepackage{epsfig}
\usepackage{amssymb}
\usepackage{amsmath}
\newif\ifpdf
\ifx\pdfoutput\undefined
\pdffalse % we are not running PDFLaTeX
\else
\pdfoutput=1 % we are running PDFLaTeX
\pdftrue
\fi

\newcommand{\bea}{\begin{eqnarray}}
\newcommand{\eea}{\end{eqnarray}}
\newcommand{\beq}{\begin{equation}}
\newcommand{\eeq}{\end{equation}}
\newcommand{\bay}{\begin{array}}
\newcommand{\eay}{\end{array}}
\newcommand{\ltw}{\tilde{\lambda}}

\begin{document}
%%%%%%%%%%%%%%%%%%%%%%%%%%%%%%%%%%%%%%%%%%
%Some more stuff to get graphics to work
\ifpdf
\DeclareGraphicsExtensions{.pdf, .jpg}
\else
\DeclareGraphicsExtensions{.eps, .jpg}
\fi
%%%%%%%%%%%%%%%%%%%%%%%%%%%%%%%%%%%%%%%%%%

\title{Shining on an Orbifold}

\author{Andrew E. Blechman}

\affiliation{Department of Physics and Astronomy,
The Johns Hopkins University, 3400 North Charles Street, Baltimore, MD 
21218}

\date{\today  \\  \vspace{1.0in} }
\vspace{2.0in}

%%%%%%%%%%%%%%%%%%%%%%%%%%%%%%%%%%%%%%%%%%%%%%%%%%%%%%%%%%%%%%%%%
\begin{abstract}
\setlength\baselineskip{18pt}

By shining a hypermultiplet from one side of the bulk of a flat five-dimensional orbifold, 
supersymmetry can be broken by boundary conditions.  
The extra dimension is stabilized in a supersymmetric way, and by computing 
the four-dimensional effective potential for the radion it is shown that 
supersymmetry breaking does not damage our radius stabilization mechanism.  The
low energy theory contains the radion and two complex scalars that are
massless in the global supersymmetric limit and are stabilized by tree level supergravity
effects.  It is shown that radion mediation can play the dominant role in
communicating supersymmetry breaking to the visible sector.
It is also shown that at tree level, contact terms are exponentially suppressed.

\end{abstract}

\maketitle

\section{Introduction}

Supersymmetry (SUSY) and extra dimensions are some of the most active areas 
of research in high energy physics today.  In addition to their 
mathematically aesthetic value, they might be able to solve the hierarchy 
problems of particle physics, and both are motivated by string 
theory.  However, the world we live in is four dimensional and not 
supersymmetric.  Therefore if SUSY exists it must be broken, probably 
spontaneously.  And if extra dimensions exist they must be compactified or 
in some way hidden.  These two constraints provide a wealth of possible 
phenomenology; see for example \cite{Ant}.

Extra dimensions have another problem.  If you naively try to compactify 
them, they are inherently unstable due to Casimir forces.  Therefore any 
self-consistent model with extra dimensions must include a way to stabilize 
the dimensions against these quantum fluctuations.

One method of doing just that is known as the Goldberger-Wise (GW) mechanism 
\cite{GW}.  This was originally designed to stabilize the extra dimension of 
the RS1 Model \cite{RS1}.  Goldberger and Wise proposed including a scalar 
field that lived in the bulk but that had independent potentials localized 
on branes at the two orbifold fixed points.  These independent potentials 
generate a profile for the scalar, and matching boundary conditions 
enforces a stabilized extra dimension.

A similar idea that involves supersymmetry was considered in \cite{AHSW}.  
In this paper, the extra dimension is a circle and a hypermultiplet has a 
source term on a brane located at $y=0$.  This hypermultiplet has an 
exponential profile in the bulk.  Then a ``probe brane'' is included that 
interacts with the hypermultiplet.  The F-flatness conditions conspire to 
stabilize the radius of the extra dimension by fitting boundary conditions.  
This method can also be used to break supersymmetry by fixing the model so 
that it is impossible to satisfy the F-flatness conditions and the boundary 
conditions at the same time.  Breaking SUSY in this way is generally called 
``shining'' \cite{AHSW,AD}.

This paper extends this idea to a flat orbifold.  A single hypermultiplet lives 
in the bulk, and it has sources on branes located at both orbifold fixed 
points.  Fitting boundary conditions overconstrains the problem and forces 
the radius to be stabilized.  A very nice side effect of this model is that 
supersymmetry need not be broken in order to stabilize the radius.
Once we stabilize the radius of the extra dimension we can break 
supersymmetry using the same technique.  We shine another 
hypermultiplet from the brane at $y=0$ and find that we cannot match 
boundary conditions and preserve supersymmetry at the same time.  We show 
that that this SUSY breaking does not have any sizeable effect on the radius 
stabilization mechanism.  This method is improved from \cite{AHSW} since the
orbifold geometry means that we do not need any chiral superfields living on one
of the branes.

Our model is similar to one proposed previously by Maru and Okada, but they
consider the warped case \cite{Nobu}.  However, they claim that there is no
viable flat space limit.  We show why this is not correct.  We will also 
correct a claim about the zero modes of the 4D effective theory.

In the next section we will present the model and show how the shining 
mechanism can be used to both stabilize the radius and break supersymmetry.  
In the following section we will consider the four-dimensional effective 
theory that reproduces the low energy physics.  We will also discuss how 
supergravity effects help stabilize the flat directions, and 
how radion \cite{CL} and anomaly \cite{RS0} mediated supersymmetry breaking 
can occur.

\section{The Model}

In this section we will present the model in terms of $N=1$ superfields in 
five dimensions.  We work with a single extra dimension compactified on a flat 
orbifold $S^1/\mathbb{Z}_2$:

\begin{equation}
ds^2=\eta_{\mu\nu}dx^\mu dx^\nu - R^2dy^2
\end{equation}

\noindent where we are using a mostly minus metric throughout this paper.  $R$ 
is the radius modulus field, or ``radion'', which 
parameterizes the size of the extra dimension, and $y\in[0,\pi]$ is an 
angular variable.  The orbifold parity defines a symmetry under the 
transformation $y\rightarrow -y$.  The five-dimensional fields in the theory 
will be either even or odd under this parity.

This model consists of two hypermultiplets $(\Phi,\Phi^c)$ and 
$(\Psi,\Psi^c)$ that are shined across the bulk from a brane located at 
$y=0$.  One of these hypermultiplets will be used to stabilize the extra 
dimension while the other one will be used to break supersymmetry.  In the 
convention that we use, the conjugate superfields are even under the orbifold 
parity while the other chiral superfields are odd.

The five-dimensional action for our model is given by \cite{MP,AGW}:

\begin{eqnarray}
S&=&\int d^4x\hspace{1mm}dy\int 
d^4\theta\hspace{1mm}\varphi^\dagger\varphi\hspace{1mm}\frac{T+T^\dagger}{2}\left\{-3M_5^3+\Phi^\dagger\Phi+\Phi^{c\dagger}\Phi^c+\Psi^\dagger\Psi+\Psi^{c\dagger}\Psi^c\right\} 
\nonumber \\
& &+\int d^4x\hspace{1mm} dy\int d^2\theta\hspace{1mm} 
\varphi^3\left\{\Phi^c(\partial_y+mT)\Phi+\Psi^c(\partial_y+\mu 
T)\Psi\right\} + {\rm h.c.} \nonumber \\
& &+\int d^4x\hspace{1mm} dy\int d^2\theta\hspace{1mm} 
\varphi^3\left\{\Phi^c[J\delta(y)-J^{'}\delta(y-\pi)]+\Psi^c 
K\delta(y)+\alpha\delta(y)\right\} + {\rm h.c.} \label{model}
\end{eqnarray}

\noindent where $\varphi$ is the conformal compensator and $T$ is the radion 
superfield\footnote{Notice the $T$ dependence in the bulk mass term for the 
hypermultiplet.  This dependence was not included in Equations 11-14 of 
\cite{MP}.  However  their later inclusion of $F_T$ in the action was correct, 
so this does not change any of their results.  Therefore we assume that this 
is simply a typo in their paper.} (see Appendix A).  $\alpha$ is a constant 
superpotential living on the $y=0$ brane that will be used to cancel the 
cosmological constant after SUSY breaking.  Notice that this action has a
$U(1)_R$ symmetry in the bulk and the $y=\pi$ brane with $R(\Psi^c)=R(\Phi^c)=
+2$ and all other superfields neutral.  This symmetry is explicitly broken on
the $y=0$ brane by the $\alpha$ term.  This will be important later.  Also 
notice that if we extend our domain 
in $y$ to the covering space $y\in[-\pi,\pi]$ the mass terms contain a sign 
function.  We leave this out to avoid the cumbersome notation, but it is 
very important when going to the four dimensional effective theory.

This model is virtually identical to the model of Maru and Okada 
\cite{Nobu}.  In that paper the authors stabilized the extra dimension in 
the case of a warped background using a hypermultiplet with delta-function
sources on both branes.  However they claim that the only way this can be done 
is in warped space and that if you take the flat space limit you get a runaway 
potential for the radion.  This is not the case if you 
take the appropriate flat space limit.  Specifically, they parameterized 
their bulk masses in terms of a c-parameter: $m=(\frac{3}{2}+c)k$ where $k$ 
is the curvature in the warp factor.  Then if you naively take the limit 
$k\rightarrow 0$ the bulk masses would vanish and the radion would no longer 
be stabilized.  The appropriate thing to do is to take the limit as 
$k\rightarrow 0$ while holding the bulk mass fixed.  It is easy to take this 
limit in their paper and we get the same results presented here for the 
radion potential.

As a first step in analyzing the model we ignore supergravity contributions, 
so $T=R$ and $\varphi=1$; in other 
words, $F_T=F_\varphi=0$.  We will come back to this in a later section.  
With these conditions the remaining F-term equations of motion are:

\begin{eqnarray}
RF_\Phi^c&=&(mR+\partial_y)\phi+\left[J\delta(y)-J^{'}\delta(y-\pi)\right] 
\\
RF_\Psi^c&=&(\mu R+\partial_y)\psi+K\delta(y) \\
RF_\Phi&=&(mR-\partial_y)\phi^{c} \\
RF_\Psi&=&(\mu R-\partial_y)\psi^{c}
\end{eqnarray}

Supersymmetry is maintained if we can find ($y$-dependent) vevs of the scalar
fields so that all of the above F-terms vanish.
Let us first consider the F-flatness condition $F_\Phi^c=0$.  The first 
delta function gives the boundary condition $\phi(0)=-\frac{J}{2}$ so there 
is a unique solution:

\begin{equation}
\phi(y)=-\frac{J}{2}\Theta(y)e^{-mR|y|}
\end{equation}

\noindent where $\Theta(y)$ is the Heaviside step function with the 
convention $\Theta(0\pm)=\Theta(\pi\mp)=\pm 1$.  The boundary condition at $y=\pi$ then 
overconstrains the problem and fixes the radius:

\begin{equation}
R=\frac{1}{m\pi}\log\left(\frac{J}{J^{'}}\right)
\label{radius}
\end{equation}

\noindent Hence this model stabilizes the size of the extra dimension as 
long as $|J|>|J^{'}|$ and they each have the same sign.

The $\Psi$ sector breaks supersymmetry through the shining mechanism 
\cite{AHSW}.  To understand how this works notice that if we set 
$F_\Psi^c=0$ we can write down the solution:

\begin{equation}
\psi(y)=-\frac{K}{2}\Theta(y)e^{-\mu R|y|}
\label{psi}
\end{equation}

\noindent The coefficient is set by the delta function source on the $y=0$ 
brane.  Notice however that there is no source on the $y=\pi$ brane; 
combined with the fact that $\psi(y)$ is an odd field the boundary condition 
is $\psi(\pi)=0$.  This boundary condition is inconsistent with Equation 
(\ref{psi}), and therefore supersymmetry is broken on the boundary at 
$y=\pi$.

Finally let us consider the last two F-terms.  Setting these equations to 
zero gives the general results:

\begin{eqnarray}
\phi^{c}&=&Be^{mR|y|} \\
\psi^{c}&=&Ce^{\mu R|y|}
\end{eqnarray}

\noindent The coefficients $B$ and $C$ are arbitrary and represent an 
indetermination of the four-dimensional zero modes of these scalars.  Hence, 
upon integrating out the fifth dimension these fields correspond to flat 
directions.

That there are two flat directions in our theory should come as no surprise
\cite{O}.  $\psi^{c}$ is the scalar field in the multiplet that breaks supersymmetry 
$(F_\Psi^c\neq 0)$, so it is expected to be flat at tree level.  That 
$\phi^c$ is also a flat direction should not surprise us either.  It is due 
to the fact that the condition $F_\Phi^c=0$ was used to stabilize the 
extra dimension, i.e.: give the radion a mass.  This leaves over an extra 
degree of freedom corresponding to the massless $\phi^c$.
This interpretation of the flat directions differs from \cite{Nobu}; this
difference will be clarified when we discuss the 4D effective theory.

\section{4D Spectrum}

Now we will consider the four-dimensional effective theory
generated by the action in Equation (\ref{model}).  In the first section we will
derive the effective potential for the radion and SUSY breaking by setting
all the hyper-scalars to their vevs from the previous section.  In the 
next section we will consider the contributions coming from the hyper-scalars
and write down an effective superpotential and Kahler potential that captures
these effects.  In the third section we will consider the lowest order effects 
of supergravity (turning $F_\varphi$ and $F_T$ back on).  In the final section
we will look at how other fields are affected by the shining field.  We consider 
the specific examples of putting matter on one of the branes, and of putting a 
gauge field in the bulk. 

\subsection{Radion Potential}

We now wish to construct an effective potential for the radion.  In the 
process we will also be able to parameterize the size of supersymmetry 
breaking.  In order to do this we need to compute the four-dimensional 
effective potential.  Ignoring any contributions from supergravity this 
potential is given by:

\begin{equation}
V=\int_0^\pi dy R\left[|F_\Psi|^2+|F_\Psi^c|^2+|F_\Phi|^2+|F_\Phi^c|^2\right]
\label{pot}
\end{equation}

\noindent There is a very nice way to understand Equation (\ref{pot}) that was 
presented in \cite{AHSW}: think of the extra dimension coordinate $y$ as a 
(continuous) index for the chiral superfields.  Then the potential is 
nothing more than the sum of all of the magnitude-squared F-terms, which is 
precisely what Equation (\ref{pot}) is.  $F_\Psi$ and $F_\Phi$ are 
proportional to the flat directions so they will not contribute to the 
effective potential at tree level.  We will see how the zero modes of the 
even scalars contribute to the effective potential in a later section.  This 
leaves two terms to calculate.

Supersymmetry is explicitly broken in the $F_\Psi^c$ term.  To isolate that 
result we must consider the full equations of motion for the scalar field upon
integrating out the auxiliary fields.  Rather than do that explicitly, we
employ the following trick, which is equivalent.  We insist that the boundary 
conditions on the fields are sacred; therefore $\psi(\pi)=0$ must be 
enforced.  We have already seen that this condition cannot be satisfied for 
$F_\Psi^c=0$ but we can get as close as possible if we make the following 
ansatz:

\begin{equation}
\psi(y)=-\frac{K}{2}\Theta(y)\left[e^{-\mu R|y|}-e^{-\mu 
R\pi}f(y)\right]
\end{equation}

\noindent where $f(y)$ is some function that satisfies the boundary 
conditions $f(0)=0$, $f(\pi)=1$.  This will enforce the boundary condition 
but at the cost of introducing a term into the potential:

\begin{equation}
\Delta V=\int_0^\pi dy\frac{K^2}{4R}e^{-2\mu R\pi}|\partial f-\mu Rf|^2
\label{deltaV}
\end{equation}

\noindent Now we can chose this function to minimize the potential.  Performing this 
minimization using variational methods and using the boundary conditions 
gives:

\begin{equation}
f(y)=\frac{\sinh(\mu Ry)}{\sinh(\mu R\pi)}
\label{f}
\end{equation}

\noindent We can plug this result back into Equation (\ref{deltaV}) and 
integrate over $y$ to get:

\begin{equation}
\Delta V=\frac{1}{2}\frac{\mu K^2}{e^{2\mu\pi R}-1}
\label{susybreak}
\end{equation}

$F^c_{\Phi}$ vanishes only when $R=r_0$, the stabilized radius defined in
Equation (\ref{radius}).  For an arbitrary radius, $F^c_{\Phi}\neq 0$ and 
we can repeat the above steps exactly for $\phi(y)$ appearing in $F_\Phi^c$.  
We try the ansatz:

\begin{equation}
\phi(y)=-\frac{J}{2}\Theta(y)\left[e^{-mR|y|}-\left(\frac{J^{'}}{J}-e^{-mR\pi}\right)g(y)\right]
\end{equation}

\noindent where $g(y)$ has the same boundary conditions as $f(y)$.  Indeed, 
upon minimizing the potential we find that $g(y)$ has the same form as 
$f(y)$ with $\mu$ replaced by $m$.  Plugging it back into Equation (\ref{pot}) 
and integrating over $y$ we find:

\begin{equation}
V(R)=\frac{1}{2}\frac{m(J-J^{'}e^{m\pi R})^2}{e^{2m\pi R}-1}+\Delta V
\label{rpot}
\end{equation}

This potential is minimized for the radius given in Equation (\ref{radius}).  
Near this stabilized radius $\Delta V\sim\mu K^2(J/J^{'})^{-2\mu/m}$ does not 
give a significant correction relative to 
the first term due to the exponential suppression for even moderate values of 
the parameters.  For concreteness, we chose the parameters: $J=K=M_5^{3/2}/10$, 
$J^{'}=M_5^{3/2}/100$, $\mu=M_5/10$ and $m=M_5/75$.  Then we find 
$R\sim 55l_5$ where $l_5$ is the 5D Planck length.  This generates a 
compactification scale $M_c\sim 0.02M_5$.  Using the well-known relation
$M_P^2=M_5^3/M_c$, we estimate $M_5\sim 10^{17}$ GeV.  We estimate the vacuum 
energy at this radius to be $M_{SUSY}=3\times 10^{-5}M_5\sim 10^{12}$ GeV.

We can take the second derivative of this potential to find the mass of the 
radion.  After taking into account the normalization of the radion (see 
Appendix A) we find $m_r\sim 10^{-3}M_P\sim 10^{15}$ GeV for the above 
values of the parameters.

\subsection{Higher Modes and the Effective Superpotential}

To get the effective scalar potential in four dimensions we must do a KK 
expansion of the fields.  The details of this are reviewed in Appendix B.  
Here we quote the results:

\begin{eqnarray}
\phi(x,y)&=&-\frac{J}{2}\Theta(y)e^{-mR|y|}+\sqrt{\frac{2}{\pi}}\sum_{n}\phi_n(x)\sin\left(ny\right)  \label{phi}  \\
\phi^c(x,y)&=&-B(x)e^{+mR|y|}+\sqrt{\frac{2}{\pi}}\sum_{n}\phi^c_n(x)\sin\left[ny+\tan^{-1}\left(\frac{n}{mR}\right)\right] 
  \label{phic}
\end{eqnarray}

\noindent and similarly for $(\psi,\psi^c)$ with $(B,m,J)\rightarrow(C,\mu,K)$.  
The KK masses are given by the simple relation: $M_n^2=m^2+n^2/R^2$ ($n>0$) 
for both $\phi$ and $\phi^c$ ($\psi$ and $\psi^c$) and $M_B=M_C=0$.  The minus 
sign in front of 
$B(x)$ is inserted for later convenience.  The first term in Equation 
(\ref{phi}) is a $y$-dependent vev.  There is no zero mode for the odd field, 
as explained in Appendix B.  This is another correction to \cite{Nobu}, who 
suggest that the zero mode of the odd field corresponds to the flat direction.  This 
role is played by the even zero mode $B(x)$, as explained earlier.

To get the 4D effective theory we insert this result into the full 
five-dimensional Lagrangian and integrate over $y$.  Since the KK modes all 
have masses at the compactification scale or higher they should not seriously
affect the low energy physics; we will see that they decouple below.  We also 
have the ($y$-dependent) vev of the odd field; that just gives us the 
potential previously calculated in Equation (\ref{rpot}).  We are left with 
the zero mode for the even field:

\begin{equation}
\mathcal{L}_4=\int_0^\pi dy\hspace{1mm}e^{2mRy}|\partial 
B|^2=\frac{1}{2m}(e^{2mR\pi}-1)|\partial B|^2 + \mathcal{O}(\partial R)
\label{unnormL}
\end{equation}

Now define $R=r_0+r$.  We can canonically normalize the field $B(x)$ by 
making the field redefinition: $B\rightarrow B\left(\frac{2m}{e^{2m\pi 
r_0}-1}\right)^{1/2}$ and we finally have (after including the 
$\psi$-sector):

\begin{eqnarray}
\mathcal{L}_4 &=& |\partial B|^2+\lambda|\partial B|^2\left[2\pi 
mr+2\pi^2m^2r^2+\cdots\right] \nonumber \\
& & +|\partial C|^2+\tilde{\lambda}|\partial C|^2\left[2\pi \mu 
r+2\pi^2\mu^2r^2+\cdots\right] + \mathcal{O}(\partial r)\nonumber \\
& & -V(r_0+r)\label{4dlag}
\end{eqnarray}

\noindent where $V(r_0+r)$ is the potential in Equation (\ref{rpot}) and the 
terms in brackets come from expanding $2e^{\pi mr}\sinh(\pi mr)$.  Using 
Equation (\ref{radius}):

\begin{eqnarray}
\lambda &=& \frac{1}{1-e^{-2m\pi r_0}}=\frac{1}{1-(J^{'}/J)^2} \\
\tilde{\lambda} &=& \frac{1}{1-e^{-2\mu\pi 
r_0}}=\frac{1}{1-(J^{'}/J)^{2\mu/m}}
\end{eqnarray}

\noindent Equation (\ref{4dlag}) is the four-dimensional effective 
Lagrangian for the canonically normalized scalar field zero modes and their 
lowest order couplings to the radion.

In addition to Equation (\ref{4dlag}), there are also terms that involve the
derivative of R.  These terms are already quadratic in the $B$ field, so they
represent other higher order effects that do not interest us here.

The higher KK modes do not have any problem or ambiguity in their coupling to
the radion, which comes from the KK mass term:

\begin{displaymath}
\Delta\mathcal{L}=-\sum_n\phi_n^{c\dagger}\left\{\partial^2+\left[m^2+\frac{n^2}{r_0^2}\left(1+\frac{r}{r_0}\right)^{-2}\right]\right\}\phi_n^c
\end{displaymath}

Now we would like to write down the four-dimensional Lagrangian in terms of 
superfields.  The only relevant fields that appear in the low energy theory are the $B,C$ scalars and the radion.  The kinetic terms and the interaction terms can be derived from a Kahler potential\footnote{There is a subtlety here.  When writing down the Kahler and superpotential we must match to the component 
Lagrangian before rescaling the fields.  So Equations (\ref{Kahler}) and 
(\ref{superpot}) are actually found from matching to Equation 
(\ref{unnormL}) after a field redefinition 
$B\rightarrow\sqrt{2m}B$, $C\rightarrow\sqrt{2\mu}C$ to get the dimensions 
right.}:

\begin{equation}
K_4=\mathcal{B}^\dagger\mathcal{B}\left(e^{m\pi(T+T^\dagger)}-1\right)+\mathcal{C}^\dagger\mathcal{C}\left(e^{\mu\pi(T+T^\dagger)}-1\right) 
  \label{Kahler}
\end{equation}

\noindent where $\mathcal{B}$ and $\mathcal{C}$ are the four dimensional 
chiral superfields containing $B$ and $C$ respectively.  We also need to 
write down a superpotential that gives us Equation (\ref{rpot}):

\begin{equation}
W_4=-\sqrt{\frac{m}{2}}\left(J-J^{'}e^{m\pi 
T}\right)\mathcal{B}-\sqrt{\frac{\mu}{2}}K\mathcal{C}
\label{superpot}
\end{equation}

\noindent This choice for the Kahler potential and superpotential will, after 
the appropriate canonical rescaling, reproduce Equation (\ref{4dlag}).

\subsection{Effects from Supergravity}

We are now in a position to incorporate effects from supergravity.  We start 
with the effective four-dimensional Lagrangian:

\begin{equation}
\mathcal{L}_4=\int d^4\theta\hspace{1mm}\varphi^\dagger\varphi 
\left\{-\frac{3}{2}M_5^3(T+T^\dagger)+K_4\right\}+\int 
d^2\theta\hspace{1mm}\varphi^3\left(W_4+\alpha\right)+{\rm h.c.} 
\label{superfieldL}
\end{equation}

\noindent where the first term is the supergravity contribution derived in 
\cite{LS} and $K_4$ and $W_4$ are given in Equation (\ref{Kahler}) and 
(\ref{superpot}) respectively.  The constant $\alpha$ is required to cancel the 
cosmological constant in order to properly normalize the gravitino mass 
\cite{Moha}.  The details of deriving Equation (\ref{superfieldL}) from 
the full 5D theory can be found in \cite{SS}.
The superpotential for $\mathcal{C}$ is reminiscent of the Polonyi model
\cite{Polonyi}.  In Polonyi models, the vev of the scalar field is pushed up to the
Planck scale.  This will happen here as well, but it does not do any damage to
our results \cite{Moroi}.

First we integrate out the auxiliary fields to get a scalar potential.  After 
rescaling the fields so they have canonical kinetic terms as in Equation
(\ref{4dlag}), we get:

\begin{equation}
V_4(B,C,R) = V(R)+\frac{2R}{3M_5^3}\left\{X_{\tilde{B}}[\tilde{B}-\langle\tilde{B}\rangle]^2
+X_{\tilde{C}}[\tilde{C}-\langle\tilde{C}\rangle]^2\right\}-U_0+\mathcal{O}(M_5^{-6})
\label{4dVeff}
\end{equation}

\noindent where $V(R)$ is the potential given in Equation (\ref{rpot}), 
$U_0=\frac{2R}{3M_5^3}(X_{\tilde{B}}\langle\tilde{B}\rangle^2+X_{\tilde{C}}
\langle\tilde{C}\rangle^2)$, and

\begin{eqnarray}
\tilde{B}&\equiv&\frac{1}{\sqrt{1+\epsilon^2}}(B+\epsilon C) \label{bt} \\
\tilde{C}&\equiv&\frac{1}{\sqrt{1+\epsilon^2}}(C-\epsilon B) \label{ct}
\end{eqnarray}

\noindent So we find that the $B$ and $C$ fields mix, but they can be redefined 
to have definite masses and vevs.  These quantities along with the mixing 
parameter $\epsilon$
are given in Appendix C.  If we remove the $C$ field (no supersymmetry 
breaking\footnote{This can be thought of as the limit $K\rightarrow 0$ since in 
that case the $\Psi$ sector would have no odd profile in the bulk.}) 
but there is still a cosmological constant (so $\alpha\neq 0$) then we find that 
$\langle B\rangle=\sqrt{\frac{m}{2}}\frac{3\alpha}{m^2\pi r_0J^{'}}$.  This is 
exactly as we expect from \cite{SS}.

All of the above masses and vevs depend on the radius, but we have fixed $R=
r_0$, the radius fixed by the $\Phi$-sector given in Equation (\ref{radius}). 
There is also mixing with the radion, and supergravity will give additional
contributions to the radion mass; this is not very important since $V(R)$
generates a radion mass just below the compactification 
scale while supergravity effects are all suppressed by powers of the Planck scale.  
So it is sufficient to fix the radion at $r_0$ since any radion mixing with the 
scalars will be very small.  This means that there are actually two sources of 
supersymmetry breaking: one source comes from the $C$ field directly 
$(F_C\neq 0)$, and another source from the fact that $R=r_0$ is not the true 
minimum of the potential in Equation (\ref{rpot}).  We claim that the second 
source of supersymmetry breaking is negligible compared to the first.  This can 
be seen by letting $R_{{\rm true vac}}=r_0+\delta$, where $\delta$ is small 
from the argument following Equation 
(\ref{rpot}).  In fact, a numerical analysis shows that for the values of 
parameters given, $\delta\sim 0$ to a very good approximation.  Therefore we need
not worry about these additional contributions.

The masses of the scalars can be computed for the values of the parameters 
mentioned below Equation (\ref{rpot}).  We find $m_{\tilde{B}}\sim 
10^{12}$ GeV, and $m_{\tilde{C}}\sim 10^7$ GeV.  Both of these 
masses are well below the compactification scale and $m_r$ as promised.  

Finally, we demand that the cosmological constant be tuned to zero.  Fixing the
radion to its classical value and the scalar fields to their vevs gives
$V(r_0)-U_0=0$.  This can easily be solved for $\alpha$; see Appendix C.

We can now use the formula to compute $\langle F_\phi\rangle$ and $\langle 
F_T\rangle$.  We find\footnote{Notice that these vevs are of the original
fields.  They can be computed by inverting Equations (\ref{bt}-\ref{ct}).}:

\begin{equation}
\langle F_\phi^{\dagger}\rangle=
\frac{\alpha}{M_5^3r_0}-\frac{\sqrt{2\mu}K(J^{'}/J)^{\mu/m}\langle 
C\rangle}{3M_5^3r_0}-\frac{\langle F_T^{\dagger}\rangle}{2r_0}
\label{fphi}
\end{equation}
\vspace{0.5cm}
\begin{equation}
\frac{\langle F_T^{\dagger}\rangle}{2r_0}=\frac{\frac{3\alpha}{2r_0} 
-\sqrt{\frac{\mu}{2}}\frac{K}{r_0}\left(1+\mu\pi\tilde{\lambda}r_0\right)(J^{'}/J)^{\mu/m}\langle C\rangle
-\sqrt{\frac{m}{2}}m\pi J^{'}\langle B\rangle}{2r_0(1-\lambda)m^2\pi^2\langle B\rangle^2
+2r_0(1-\tilde{\lambda})\mu^2\pi^2\langle C\rangle^2+\frac{3}{2}M_5^3}
\label{ft}
\end{equation}
\vspace{0.25cm}

\noindent The first term in Equation (\ref{fphi}) cancels the cosmological
constant; the second term comes from the SUSY-breaking $F$-term ($F_C$); the
final term is the radion-mediated contribution given in Equation (\ref{ft}).  
For the given parameters this generates $\frac{|\langle F_T\rangle|}{2r_0}\sim 
10^6$ GeV and $m_{3/2}=\langle F_\varphi\rangle\sim 10^{6}$ GeV.  These 
quantities are 
the same order of magnitude due to the large Polonyi vev $\langle C\rangle$ 
which can cancel the cosmological constant term in Equation (\ref{fphi}).  
Notice that in the limit considered earlier where $C\equiv 0$ but there is 
still a cosmological constant, we find to this order after plugging in our
result for $\langle B\rangle$ given below Equation (\ref{ct}) that 
$\langle F_T^{\dagger}\rangle=0$ and $\langle F_\varphi^{\dagger}\rangle=
\frac{\alpha}{M_5^3 r_0}$, again in agreement with \cite{SS}.

\subsection{Soft Masses from the Shining Sector}

We now ask what happens to the MSSM in our model of SUSY breaking.  In the full
5D theory, supersymmetry is broken near the brane at $y=\pi$.  Thus, we can 
place the MSSM on the brane at $y=0$ and ask if this will generate any contact
interactions in the 4D effective theory.  Such terms would look like:

\begin{equation}
\mathcal{L}_c=\int_0^\pi dy\delta(y)\int d^4\theta
\frac{Q^{\dagger}Q\Psi^{c\dagger}\Psi^{c}}{M_5^3}
\end{equation}

\noindent where $Q$ is a chiral superfield in the MSSM.

Now it is sufficient to only consider the zero mode of the hyper-scalar since
all of the KK modes have masses of order the compactification scale or higher,
and these will generate Planck and Yukawa suppressed interactions.  In this case:

\begin{equation}
\Psi^c(x,y)=\sqrt{\frac{2\mu}{e^{2\mu\pi R}-1}}\mathcal{C}(x)e^{\mu R|y|}
\label{canon}
\end{equation}

\noindent is the canonically normalized mode.  This will generate contact terms
of the form:

\begin{equation}
\mathcal{L}_c=\int d^4\theta\frac{\mu}{M_5^3}Q^{\dagger}Q\mathcal{C}^{\dagger}
\mathcal{C}e^{-2\mu\pi r_0}
\end{equation}

\noindent and this gives a contribution to the masses of the MSSM scalars:

\begin{equation}
\Delta m_{\tilde{q}}^2=\frac{\mu |F_C|^2}{M_5^3}e^{-2\mu\pi
r_0}\sim\frac{\mu M_{SUSY}^4}{M_5^3}\left(\frac{J}{J^{'}}\right)^{-2\mu/m}
\label{msquark}
\end{equation}

So these contact interactions will be exponentially suppressed at tree level. 
One could have guessed that this would be the case, since the wavefunction of
the zero mode of the even field is an exponentially increasing function of $y$. 
Thus the bulk scalar likes to spend all of its time far away from the visible
brane at $y=0$.  However, we generally expect that radiative corrections might 
spoil this result and must be checked in models that incorporate this shining 
mechanism.

Now consider putting a gauge field in the bulk (for simplicity, let it be a 
$U(1)$ gauge field, but it does not have to be).  This would give an extra
contribution to the action:

\begin{equation}
\Delta\mathcal{L}_4=\int d^2\theta \frac{T}{4g_5^2}\mathcal{W}_\alpha
\mathcal{W}^\alpha\vspace{0.25in}+{\rm h.c.}
\label{rmsb}
\end{equation}

\noindent This term generates a contribution to the gaugino mass through radion 
mediation \cite{CL}:

\begin{equation}
\Delta m_{1/2}^{(RMSB)}=\frac{\langle F_T\rangle}{2r_0}\sim m_{3/2}
\end{equation}

Anomaly mediation also gives a contribution to the gaugino 
masses.  This formula is complicated somewhat by the fact that the Polonyi model
has a Planck-scale vev \cite{BPM}, but the important point is that
$\Delta m_{1/2}^{(AMSB)}\ll \Delta m_{1/2}^{(RMSB)}$ due to a loop 
factor.  So radion mediation is the dominant contribution to $m_{1/2}$ coming
from supergravity.

We can also have contact interactions between the gauge field and the 
shining field:

\begin{equation}
\Delta\mathcal{L}=\int_0^\pi dy\int d^2\theta
\frac{\Psi^{c}\mathcal{W}_\alpha\mathcal{W}^\alpha}{M_5^{3/2}}
\label{cww}
\end{equation}

After plugging in Equation (\ref{canon}), this will introduce a new 
contribution to the gaugino mass:

\begin{equation}
\Delta m_{1/2}^{(C)}=\frac{|F^C|}{M_5}\sim\frac{M_{SUSY}^2}{M_5}
\label{cm12}
\end{equation}

\noindent Thus this contact term gives a contribution to the gaugino mass 
$\Delta m_{1/2}^{(C)}\sim 10^7$ GeV, which is comparable to 
$\Delta m_{1/2}^{(RMSB)}$ at tree level.

We can suppress this contribution to the gaugino mass by making use of the
$U(1)_R$ symmetry mentioned below Equation (\ref{model}).  From Equation
(\ref{rmsb}) we see that $R(\mathcal{W}_\alpha)=+1$ so that the contact term in
Equation (\ref{cww}) breaks the R symmetry by 2 units.  Thus it can only be
generated on the $y=0$ brane where the $\alpha$ term has already broken the
R-symmetry.  Thus the generated contact term in Equation (\ref{cww}) will come 
with a delta function.

\[
\Delta\mathcal{L}=\int_0^\pi dy\delta(y)\int d^2\theta
\frac{\Psi^{c}\mathcal{W}_\alpha\mathcal{W}^\alpha}{M_5^{3/2}}
\]

\noindent Plugging in Equation (\ref{canon}) for $\Psi^c$ and integrating over
$y$ will now generate an exponentially suppressed contribution to the mass in
analogy with Equation (\ref{msquark}):

\[m_{1/2}^{(C)}\sim\frac{M_{SUSY}^2}{M_5}\left(\frac{J}{J^{'}}\right)^{-\mu/m}
\ll \Delta m_{1/2}^{(RMSB)}\]

So we find that it is possible for the RMSB contribution to dominate the gaugino 
mass.

\section{Discussion}

This paper has extended the shining mechanism of supersymmetry breaking to the
geometry of flat orbifolds.  This is a very nice way to break supersymmetry via 
a hidden sector in extra dimensions.  It avoids the need for extra
superfields living on the boundary branes as in \cite{AHSW}.  It can easily be
extended to other interesting situations such as matter or gauge fields in the 
bulk, where radion mediation can play an important role.

This paper has also clarified some of the issues raised in \cite{Nobu}.  In
particular, contrary to their claim, it {\it is} possible to fit their model to
the flat case and there is nothing special about the warped geometry.  We have
also clarified the role of the zero modes in the low energy theory.

In addition we have shown how supergravity plays the usual role of radiative
corrections in stabilizing the flat scalars.  This is because our model is
actually a Polonyi model in disguise, which is a free field theory in the limit
$M_5\rightarrow\infty$.

This model of supersymmetry breaking only introduces exponentially
suppressed contact terms at tree level when the MSSM is put on the 
brane at $y=0$.  So it might be possible to generate realistic soft masses for 
the squarks and sleptons.
In addition, radion mediated SUSY breaking might play an important part if the
bulk contact terms can be suppressed.  Here, this was accomplished by imposing an
R-symmetry that originally appeared as an accidental symmetry in the bulk and is
broken on the brane at $y=0$.

The classic example of a model with radion mediation as the dominant mechanism
of SUSY breaking is the ``no-scale model" \cite{NS}, where $F_\varphi\equiv 0$
\cite{MP}.  This model is known to be unstable after radiative corrections are
included.  Recently, it has been improved by including a general stabilization
mechanism and a constraint was derived to keep the model ``almost no-scale" 
\cite{ANS}:

\[ \langle K_{T^\dagger T}\rangle\ll\frac{M_5^3}{2\pi r_0} \]

\noindent where $K$ is a radius-stabilizing Kahler potential.
This constraint corresponds to making sure that $F_\varphi$ remains
small relative to $F_T/r_0$.  The model considered here violates this 
constraint: both sides of the innequality are the same order of magnitude.  
This is because our model has $F_\varphi\sim F_T/r_0$.  
Anomaly mediation is then suppressed by a loop factor, not a small $F_\varphi$. 
This is what leads to dominant radion mediation.

Finally, notice that this model, although in flat space, has a Kahler potential
that depends on the exponential of the radion.  This is reminiscent of warped
space, and there might be a corresponding reinterpretation of the effective
four-dimensional theory.  This could lead to interesting consequences for
AdS/CFT, warped supergravity, etc, and is left for future research.

\vspace{0.5in}
\noindent{\bf Acknowledgments}
\vspace{0.25in}

It is a pleasure to thank Paddy Fox and David Kaplan for suggesting this 
project, and especially to the latter for many useful discussions.  I also 
wish to thank Kaustubh Agashe, Jon Bagger, Roberto Contino, Minho Son, Raman 
Sundrum and Chi Xiong for help on the technical points as well as for some very 
stimulating discussion.

\appendix

\section{The Radion Modulus Field}

The radion modulus field comes from the gravitational part of the action.  
To see how this comes about consider the usual Einstein-Hilbert action in 
five dimensions:

\begin{equation}
S_5=\int d^5x 
\sqrt{-G}\left\{M_5^3\mathcal{R}_{(5)}+\mathcal{L}_{(5)M}\right\}
\label{EH5}
\end{equation}

\noindent where $M_5^3$ is the five-dimensional Planck scale, $G$ is the
determinant of the 
five-dimensional metric, $\mathcal{R}_{(5)}$ is the five-dimensional Ricci 
scalar and $\mathcal{L}_{(5)M}$ contains any other fields.  We can work in 
the gauge (coordinate system) where $G_{5\mu}\equiv 0$ so the differential 
line element is:

\begin{equation}
ds^2=G_{MN}dx^Mdx^N=g_{\mu\nu}(x,y)dx^\mu dx^\nu - r^2(x)dy^2
\label{ds2}
\end{equation}

\noindent Our convention is that the metric is mostly minus.  $g_{\mu\nu}$ 
is the induced four-dimensional metric which is generally a function of the 
five-dimensional spacetime, and $G_{55}\equiv -r^2$ is assumed to be 
independent of the extra dimension.  Then $\sqrt{-G}=\sqrt{-g}\times r$ and 
upon carefully expanding the Ricci scalar, our action is:

\begin{equation}
S_5=\int d^5x 
\sqrt{-g}\left\{rM_5^3(\tilde{\mathcal{R}}_{(4)}+\delta\mathcal{R})+r\mathcal{L}_{(5)M}\right\}
\label{EH54}
\end{equation}

\noindent where
\[\mathcal{R}_{(4)}\equiv\int dy\tilde{\mathcal{R}}_{(4)}\]
is the four-dimensional Ricci scalar and $\delta\mathcal{R}[g,r]$ are the 
terms in the five-dimensional Ricci scalar that depend on the fifth 
dimension explicitly.

In a flat extra dimension the four-dimensional graviton $g_{\mu\nu}$ is 
independent of $y$, so the $y$-dependence has been completely isolated and 
we can easily perform the integral over the fifth 
dimension\footnote{It is not this simple in general.  For example, in the RS 
model there is also warp factor and more work needs to be done.  However, it 
is not much harder to handle this case.}.  However, the graviton kinetic 
term is no longer canonical.  To fix this problem we can do a Weyl rescaling 
of the metric \cite{Wald}:

\begin{eqnarray*}
g_{\mu\nu}&\longrightarrow&\Omega^2g_{\mu\nu}\\
g^{\mu\nu}&\longrightarrow&\Omega^{-2}g^{\mu\nu}
\end{eqnarray*}

\noindent Under this transformation:

\begin{eqnarray}
\sqrt{-g}&\longrightarrow&\Omega^4\sqrt{-g} \\
\mathcal{R}_{(4)}&\longrightarrow&\Omega^{-2}\left\{\mathcal{R}_{(4)}+6\left[(\partial(\log\Omega))^2+\partial^2(\log\Omega)\right]\right\} 
  \label{RicciT}
\end{eqnarray}

It is clear from these equations that in a flat extra dimension 
$S^1/\mathbb{Z}_2$ where $y\in[0,\pi]$:

\begin{equation}
\Omega^2=\frac{M_4^2}{\pi rM_5^3}
\end{equation}

\noindent will generate the canonical kinetic term for the four-dimensional 
graviton, where $M_4$ is the usual 4D Planck scale.  In addition it will also 
generate a canonical kinetic term for the radion:

\begin{equation}
S_4=\int d^4x 
\sqrt{-g}\left\{M_4^2\mathcal{R}_{(4)}+\frac{1}{2}(\partial\rho)^2+\cdots\right\}
\end{equation}

\noindent where I have defined the canonical radion field:

\begin{equation}
\rho\equiv\sqrt{12}M_4\log\left(\frac{r}{r_0}\right)
\end{equation}

\noindent where $r_0$ is the classical radius, assumed to be stabilized.  
Notice that the canonical radion of flat space is the logarithm of R, as 
opposed to the exponential of R in the warped case \cite{GW2}.

By letting $r=r_0+\delta r$ and expanding the logarithm the 
canonical radion field is related to the usual radion modulus by a constant:

\begin{equation}
\rho=\frac{\sqrt{12}M_4}{r_0}\delta r +\mathcal{O}\left(\left(\delta 
r/r_0\right)^2\right)
\label{rho}
\end{equation}

\noindent So to compute the radion mass in Planck units ($M_4\equiv 1$) we 
compute the radion potential at quadratic order: 
$V(r)=\frac{1}{2}\mu_r^2(\delta r)^2$.  Then $M_{{\rm radion}}=\mu_r 
r_0/\sqrt{12}$.

Using what we now know it is easy to see how the radion can be incorporated 
into the linearized supergravity action by extending into 
superspace.  To see how this is done notice that the radion modulus appears in 
a $N=1$ chiral superfield that contains the $\mathbb{Z}_2$-even fifth 
components of the fields that appear in the 5D supergravity multiplet.  This 
field is called the ``radion superfield'':

\begin{equation}
T(x,\theta)=(r+iB_5)+\sqrt{2}\theta\Psi_R^5+\theta^2F_T
\end{equation}

\noindent where $B_5$ is the fifth component of the graviphoton and 
$\Psi_R^5$ is the fifth component of the right-handed gravitino.  This is 
derived in many places such as \cite{LLP}.  Now all we have to do is to 
include the radion superfield everywhere that it should appear so that we 
reproduce the correct action in terms of component fields.  This was done in 
\cite{MP,AGW} for a general class of theories \cite{T-error}.  It is important 
to notice that this matching must be done before the Weyl rescaling, as 
explained in \cite{LS}.

\section{KK Decomposition}

In this appendix we will derive the KK decomposition for our model.  For 
simplicity the analysis will only be done for the $\Phi$-sector.  It is 
exactly the same for the $\Psi$-sector.

To perform the decomposition it is necessary to write out the Lagrangian for 
the scalar components by integrating out the auxiliary fields from Equation 
(\ref{model}).  This is given by\footnote{Recall: $\partial_5\equiv\partial_y/R$}:

\begin{eqnarray}
\mathcal{L}&=&\phi^{c\dagger}\left[-\partial^2+\partial_5^2-m^2-\frac{2m}{R}\left(\delta(y)-\delta(y-\pi)\right)\right]\phi^c 
\nonumber \\
& &+\phi^\dagger\left[-\partial^2+\partial_5^2-m^2+\frac{2m}{R}\left(\delta(y)-\delta(y-\pi)\right)\right]\phi 
\nonumber \\
& &-\phi\left(-\partial_5+m\Theta(y)\right)S-\phi^\dagger\left(-\partial_5+m\Theta(y)\right)S
\label{scala}
\end{eqnarray}

\noindent where $S=J\delta(y)-J^{'}\delta(y-\pi)$.  The extra delta 
functions in each bracket come from the fact that the mass term is an odd term.  This yields 
the equations of motion:

\begin{eqnarray}
\left[-\partial^2+\partial_5^2-m^2-\frac{2m}{R}\left(\delta(y)-\delta(y-\pi)\right)\right]\phi^c&=&0 
  \label{phiceq} \\
\left[-\partial^2+\partial_5^2-m^2+\frac{2m}{R}\left(\delta(y)-\delta(y-\pi)\right)\right]\phi&=&\left(-\partial_5+m\Theta(y)\right)S 
  \label{phieq}
\end{eqnarray}

We wish to decompose the fifth dimension so we let:

\begin{eqnarray}
\phi(x,y)&=&-\frac{J}{2}\Theta(y)e^{-mR|y|}+\sum_\lambda\phi_\lambda(x)\xi_\lambda(y) 
\\
\phi^c(x,y)&=&-B(x)e^{+mR|y|}+\sum_\lambda\phi^c_\lambda(x)\xi^c_\lambda(y)
\end{eqnarray}

\noindent where the first term in $\phi$ is the particular solution to Equation
(\ref{phieq}); it plays the role of a 
$y$-dependent vev.  This immediately takes care of the source terms for the 
$\phi$ field.  Notice that this first term is not a zero mode; the coefficient
is fixed by the inhomogeneous source terms on the right-hand side of
Equation (\ref{phieq}) which eliminate it as a degree of freedom.  The even field 
however does contain a zero mode.  We explicitly include a minus sign so that both 
$\phi$ and $\phi^c$ have the same sign in the physical region.  This is done
purely for convenience and does not change any results.

Now the equations of motion for the KK basis states are:

\begin{eqnarray}
\left[\partial_5^2-\frac{2m}{R}\left(\delta(y)-\delta(y-\pi)\right)\right]\xi_\lambda^c&=&-\lambda^2\xi_\lambda^c 
\\
\partial_5^2\xi_\lambda&=&-\lambda^2\xi_\lambda
\end{eqnarray}

\noindent where we have dropped the delta functions in the equation for $\xi_\lambda$ 
since it is an odd field and therefore does not feel the delta functions on 
the boundary.  Then $\phi_n(x),\phi_n^c(x)$ are the KK modes with masses 
$M_\lambda^2=m^2+\lambda^2$.

The equation for $\xi(y)$ is a very easy equation to solve.  Remembering 
that the odd fields must vanish at the boundaries:

\begin{equation}
\xi_\lambda(y)=\sqrt{\frac{2}{\pi}}\sin\left(ny\right)\hspace{1in}\lambda=\frac{n}{R}
\end{equation}

\noindent The equation for $\xi^c$ is not any more difficult.  It is just 
the Schrodinger equation with delta function potentials and symmetric 
boundary conditions.  We find that $\lambda^2<0$ cannot happen so there are 
no ``bound states''.  The final solution is:

\begin{equation}
\xi_\lambda^c(y)=\sqrt{\frac{2}{\pi}}\sin\left[ny-\tan^{-1}\left(\frac{n}{mR}\right)\right]\hspace{1in}\lambda=\frac{n}{R}
\end{equation}

\noindent These are the modes that appear in Equation 
(\ref{phi}-\ref{phic}).  They have been normalized so that $\int_0^{\pi} 
dy\hspace{1mm}\xi_\lambda\xi_{\lambda^{'}}=\delta_{\lambda\lambda^{'}}$.  
Also notice that the zero mode of $\phi^c$ is orthogonal to the higher 
modes, which is easily checked.

\section{Supergravity Contributions}
In this section, we present the masses and vevs of the hypermultiplets after the
lowest-order supergravity effects are taken into account.  We make the following
definitions:

\begin{eqnarray}
a&=&\frac{1}{2}m^3\pi^2 J^{'2}  \\
b&=&\ltw\pi\mu^2K^2(J^{'}/J)^{2\mu/m}\left(\frac{2}{r_0}+\frac{1}{2}\mu\pi\ltw\right)  \\
d&=&(m\mu)^{3/2}JK(J^{'}/J)^{1+\mu/m}\left(\ltw+\frac{2\pi}{\mu r_0}\right) \\
f&=&\frac{6\pi}{r_0\sqrt{2}}m^{3/2}J^{'}  \\
g&=&\frac{6\pi}{r_0\sqrt{2}}\mu^{3/2}(J^{'}/J)^{\mu/m}\ltw K  
\end{eqnarray}

\noindent These parameters are defined up to terms with $R\neq r_0$.  Then in 
terms of these parameters, the masses, vevs and mixing 
parameter in the paper are:

\begin{eqnarray}
m_{\tilde{B}}^2=\frac{2R}{3M_5^3}X_{\tilde{B}}&=&\frac{r_0}{3M_5^3}\left((a+b)+\sqrt{(a-b)^2-d^2}\right) \\
m_{\tilde{C}}^2=\frac{2R}{3M_5^3}X_{\tilde{C}}&=&\frac{r_0}{3M_5^3}\left((a+b)-\sqrt{(a-b)^2-d^2}\right) \\
\langle\tilde{B}\rangle&=&\frac{\alpha}{\sqrt{1+\epsilon^2}}\cdot\frac{f+\epsilon
g}{(a+b)+\sqrt{(a-b)^2-d^2}} \\
\langle\tilde{C}\rangle&=&\frac{\alpha}{\sqrt{1+\epsilon^2}}\cdot\frac{g-\epsilon
f}{(a+b)-\sqrt{(a-b)^2-d^2}} \\
\epsilon&=&\frac{b-a}{d}+\sqrt{\left(\frac{b-a}{d}\right)^2-1}
\end{eqnarray}

\noindent where $\alpha$ is the superpotential parameter that cancels the 
cosmological constant as explained in the paper:

\begin{equation}
\alpha=\sqrt{\frac{1}{u_0}\cdot\frac{\mu K^2}{e^{2\mu\pi r_0}-1}}
\end{equation}

\noindent where $U_0=u_0\alpha^2$ is defined in the text below Equation 
(\ref{4dVeff})\footnote{$U_0$ is quadratic in $\alpha$, so $u_0$ is independent
of $\alpha$.}.

Notice that $m_{\tilde{B}}^2,m_{\tilde{C}}^2>0$ for any value of the parameters, 
so the theory is stable.


\begin{thebibliography}{99}
\bibitem{T-error} This actually only works if you ignore the gravitational
dynamics.  For instance, the action in Eq (\ref{model}) includes the term 
$\phi\partial_\mu\phi\partial^\mu B_5$ + h.c., which is not 5D Lorentz 
invariant or gauge invariant.  In this paper, however, we are only dealing with
constant fields and therefore this error is not a problem.  I thank M. Son and
M. Luty for bringing this to my attention.
\end{thebibliography}
\end{document}